\documentclass{an}
\usepackage{graphicx}
\usepackage{times}
\usepackage{fancyhdr}
\begin{document}

\title{
Evolution of Magnetic Fields in Stars Across the Upper Main Sequence: \\
I. A catalogue of magnetic field measurements with FORS\,1 at the VLT\thanks{
Based on observations obtained at the European Southern Observatory, Paranal, Chile
(ESO programmes 71.D-0308(A), 072.D-0377(A), 073.D-0464(A), and 075.D-0295(A)).
} 
}

\author{
S. Hubrig\inst{1}
\and  P. North\inst{2}
\and  M. Sch\"oller\inst{1}
\and  G. Mathys\inst{1}
}
\institute{
European Southern Observatory, Casilla 19001, Santiago 19, Chile
\and
Laboratoire d'Astrophysique, Ecole Polytechnique F\'ed\'erale
de Lausanne, Observatoire,
CH-1290~Sauverny, Switzerland
}

\date{Received $<$date$>$; 
accepted $<$date$>$;
published online $<$date$>$}

\abstract{
To properly understand the physics of Ap and Bp stars it is particularly 
important to identify the origin of their magnetic fields. For that, an
accurate knowledge of the evolutionary state of stars
that have a measured magnetic field is an important diagnostic.
Previous results based on a small and possibly biased sample suggest that the
distribution of magnetic stars with mass below 3\,$M_\odot$ in the H-R diagram
differs from that of normal stars in the same mass range (Hubrig et al. 2000).
In contrast,  higher mass magnetic Bp stars may well occupy the 
whole main-sequence width (Hubrig, Sch\"oller \& North 2005b).
In order to rediscuss the evolutionary state of upper main sequence magnetic
stars, we define a larger and bias-free sample of Ap and Bp stars with accurate
Hipparcos parallaxes and reliably determined longitudinal magnetic fields. We
used FORS\,1 at the VLT in its spectropolarimetric mode to measure the magnetic
field in chemically peculiar stars where it was unknown or poorly known as yet.
In this first paper we present our results of the mean longitudinal magnetic
field measurements in 136 stars. Our sample consists of 105 Ap and Bp stars, two
PGa stars, 17 HgMn stars, three normal stars, and nine SPB stars. A magnetic
field was for the first time detected in 57 Ap and Bp stars, in four HgMn stars,
one PGa star, one normal B-type star and four SPB stars.
\keywords{stars:chemically peculiar - stars:evolution - stars:magnetic fields}
}

\correspondence{shubrig@eso.org}

\maketitle

\section{Introduction}

Ap and Bp stars 
are main-sequence A and B stars in the spectra of which the lines of some
elements are abnormally strong (e.g., Si, Sr, rare earths) or weak (in 
particular, He). They undergo periodic variations of magnitude (in various
photometric bands) and spectral line equivalent widths; the known periods
of variability range from half a day to several decades. 
Among Ap stars, the magnetic chemically peculiar stars are
especially important.
For a long time, Ap stars were the only non-degenerate stars besides 
the sun in which direct detections of magnetic fields
had been achieved. Today, they still represent a major fraction of the known
magnetic stars. These stars generally have large-scale organized magnetic 
fields that can be diagnosed through observations of circular polarization in 
spectral lines. The unique large-scale organization of the magnetic fields in 
these stars, which in many cases appears to occur essentially under the form 
of a single large dipole located close to the centre of the star, contrasts 
with the magnetic field of late-type stars, which is most probably subdivided 
in a large number of small dipolar elements scattered across the stellar 
surface.
The fact that magnetic fields of Ap stars are more readily observable than
those of any other type of non-degenerate stars makes them a privileged
laboratory for the study of phenomena related to stellar magnetism.

To properly understand the physics of Ap stars it is particularly important to 
know the origin of magnetic fields in these stars.
It is the subject of a long debate, which is far from being closed
(e.g., Braithwaite \& Spruit 2004).
After the discovery of magnetic fields in Ap and Bp stars it was proposed that 
these stars have acquired their field at the time of
their formation or early in their evolution (what is currently observed
is then a fossil field). An alternative suggestion is that magnetic fields 
are generated and maintained by a contemporary dynamo at work
inside the star. 
Whether the A and B stars become magnetic at a certain evolutionary 
state before reaching the zero-age main sequence (ZAMS), or during the core 
hydrogen burning, or at the end
of their main-sequence life requires systematical studies of established
cluster members, binary systems, and field stars with accurate 
Hipparcos parallaxes. Until now, there is only one case of a
strongly magnetic Ap star of mass below 3\,$M_\odot$
(either member of a nearby moving cluster or supercluster or 
belonging to a binary system) which is not much evolved away from the ZAMS:
HD 66318 in NGC 2516 is claimed to have fulfilled only 16\% of its
main sequence lifetime
(Bagnulo et al. 2003; see also
Hubrig \& Schwan 1991; Hubrig \& Mathys 1994; Wade et~al.\ 1996).
Only a few double-lined spectroscopic binary systems containing a magnetic 
Ap star are currently known, and as far as the membership of Ap stars in 
distant open clusters is concerned,
we should keep in mind that such studies are mostly based upon photometry
and upon radial velocity determinations. But photometric criteria of
cluster membership are more delicate to apply to peculiar stars,
since strong backwarming effects lead to an anomalous energy distribution, thus
affecting the position of such stars in colour-magnitude diagrams. 

In our previous study of the evolutionary
state of magnetic Ap stars with accurate Hipparcos parallaxes and accurate 
measurements of the mean magnetic field modulus and mean
quadratic magnetic fields, we showed that the distribution of magnetic stars
of mass below 3\,$M_\odot$ differs  from
that of normal stars in the same
temperature range at a high level of significance (Hubrig, North \&
Mathys 2000).  
Normal A stars occupy
the whole width of the main sequence, without a gap, whereas magnetic stars are 
concentrated towards the centre of the main-sequence band. In particular, it
was found that magnetic fields appear in stars that have already
completed at least approximately 30\% of their main-sequence lifetime.

Knowing the position of the magnetic stars in the H-R diagram, it
became also possible 
to probe the evolution of magnetic field strength across the main sequence.
However, no clear picture emerged from our data.
Yet, the whole sample under study contained only 33 magnetic stars of mass
below 3\,$M_\odot$.
We exclusively selected stars for which a strong surface 
magnetic field had been definitely detected.
For these stars the mean 
magnetic field modulus, which is the average over the stellar 
disk of the modulus of the magnetic vector, has been derived through measuring 
the wavelength separation of resolved magnetically split components of spectral lines.
The mean quadratic field has been diagnosed 
from the consideration of the differential magnetic broadening of spectral lines. 
A bias was present due to the 
fact that our sample contained a large fraction (about 2/3)
of stars with rotational periods longer than 10~days, while the majority of the periods of 
magnetic stars fall between 2 and 4~days.
Clearly, there was a need for more magnetic field measurements of Ap stars
for which accurate Hipparcos parallaxes were obtained. 
To this purpose, we started a few years ago a long-term systematical search for
magnetic fields in about 100 upper main sequence chemically peculiar stars with good 
Hipparcos parallaxes. These stars were chosen in a wider range of masses, among
those whose magnetic field has been never or only poorly studied before, and
presenting a distribution 
of rotational periods more representative of that of all Ap and Bp stars. 

In this first paper, we present results of magnetic field measurements
in 136 A and B stars. The detailed analysis
of the evolution of the magnetic field across the H-R diagram in stars of different mass 
will be presented in a second paper. Some preliminary results of the analysis based 
on the magnetic field measurements from the first release of data for our 
ESO observing program have already 
been reported at various meetings (Hubrig, Sch\"oller \& North 2005b; Hubrig, North \& Szeifert 2006).
In general, we could confirm our previous results obtained from the study of Ap and Bp stars
with accurate measurements of the mean magnetic field modulus and mean
quadratic magnetic fields, i.e., that magnetic stars of mass below 3\,$M_\odot$ are concentrated 
towards the centre of the main-sequence band.
We could also show that, in contrast,  higher mass 
magnetic Bp stars may well occupy the whole main-sequence width.

\section{Basic data}

The General Catalogue of Ap and Am stars (Renson, Gerbaldi \& Catalano 1991) includes
2875 Ap stars showing abnormal enhancement of one or several elements in their
atmosphere. 
Hipparcos parallaxes have been measured for about 940 Ap stars.
371 of them have a low parallax error of $\sigma(\pi)/\pi<0.2$ 
(Gomez et~al.\ 1998).

Most studies of magnetic fields of Ap stars are based on measurements of the
mean longitudinal magnetic field, which is an average over the visible stellar
hemisphere of the component of the magnetic vector along the line of sight. It
is derived from measurements of wavelength shifts of spectral lines between
right and left circular polarization.
Before our study, only 195 Ap stars had reliably measured longitudinal fields,
ranging from tens of Gauss to about 20\,kG (Bychkov, Bychkova \& Madej 2003).
But only for 114 stars with measured magnetic fields the parallax error was
less than 20\%.
A part of these stars have been used for our study of the evolutionary state 
of magnetic stars six years ago.

For 49 stars the mean magnetic field modulus has been
derived from measurements of the wavelength separation of resolved magnetically
split components of spectral lines. The resolution of individual line
components requires a combination of sufficient magnetic field strength and
small enough projected rotational velocity.
The mean field modulus is, by definition,
much less aspect-dependent than the longitudinal field and, thus, it 
characterizes much better the intrinsic stellar magnetic field.
Unfortunately, it can only be measured in a small fraction of Ap stars that have
magnetically resolved lines.
Therefore, longitudinal field measurements represent the standard method for 
searching magnetic fields in different types of stars, and 
longitudinal field measurements, due to
their sensitivity to aspect, represent essential constraints for all
models of the geometry and the detailed structure of the magnetic
fields of these stars.
This underscores the important role of these data in understanding
magnetism in upper-main sequence stars. 

As Bagnulo et~al.\ (2002) and Hubrig et~al.\ (2004) have demonstrated,
low resolution spectropolarimetry in H Balmer lines obtained with FORS\,1
represents a powerful diagnostic method for the detection of stellar magnetic
fields. FORS\,1 is a multi-mode instrument which is equipped with 
polarization analyzing optics comprising super-achromatic half-wave and 
quarter-wave phase 
retarder plates, and a Wollaston prism with a beam divergence of 22$\arcsec$ 
in standard resolution mode.
In our latest study of magnetic fields in rapidly oscillating Ap stars 
with FORS\,1 in spectropolarimetric mode, using GRISM 600B and an 0$\farcs$4
slit,
a formal uncertainty as small as 50~G has been achieved, suggesting that the
potential of FORS\,1 for measuring magnetic fields is even higher than indicated before (Hubrig 
et~al.\ 2004). For the major part of our stellar sample we used the GRISM\,600B in the 
wavelength range 3480--5890\,\AA{} at a spectral resolution of R$\sim$2000
to cover all hydrogen Balmer lines from H$_\beta$ to the Balmer jump.
During the last semester in 2005 we used the
GRISM\,1200g to cover the H Balmer lines from H$_\beta$ to H$_8$,
and the narrowest available slit width of 0$\farcs$4 to obtain a
spectral resolving power of R$\sim$4000. The determination of the mean longitudinal fields
using FORS\,1 is described in detail in Hubrig et~al.\ (2004).
All longitudinal field determinations for the 136 stars in our sample were
obtained from observations with FORS\,1 at the VLT executed in service
mode from April 2003 to September 2005. 

In order to be able to determine the location of the observed stars in the H-R diagram,
we selected only stars for which distance and photometry can be obtained with small error
bars.
127 out of 136 stars in our sample have accurately determined
Hipparcos parallaxes with $\sigma(\pi)/\pi<0.2$.
For all objects, there exists either Geneva or Str\"omgren photometry.

Nine stars in our sample are known members
of nearby open clusters of different ages and have very accurate Hipparcos parallaxes.
Their membership has been confirmed on 
photometric, proper motion
and radial velocity grounds. They are excellent candidates for our study and the 
measurements of their magnetic fields allow us to put more 
stringent constraints on the origin of the magnetic fields.

Magnetic fields play an important role 
in the theoretical interpretation of the pulsations in rapidly oscillating Ap (roAp) stars.
However, until now, the only systematic attempt to detect and to study
their field has been done by Mathys (2003).
Still, the knowledge of the magnetic fields in many roAp stars is   
very incomplete.
Therefore, a few roAp stars, for which no magnetic field 
measurements have been reported before, have been included in our sample.

The presence of magnetic fields in so-called ``non-magnetic'' stars with
HgMn or PGa peculiarity is still a subject of debate between various observers.
To understand the role that magnetic fields 
play for the origin of chemical peculiarities in these stars,
magnetic field measurement have been carried out for 17 HgMn and two PGa stars. 

Recently, Neiner et~al.\ (2003) presented the first detection
of a magnetic field in the SPB star $\zeta$\,Cas.
It is difficult to explain why chemically peculiar hot Bp stars
and Slowly Pulsating B (SPB) stars co-exist at the same position in the H-R diagram, namely
in the SPB instability strip. The pulsation periods of SPB stars range from
about 1 to 3 days. It is especially intriguing that the magnetic fields 
of hot Bp stars do not show any detectable variations or vary with periods
close to 1 day. A small sample of SPB stars and a few monoperiodic B stars with a 
non-homogeneous distribution of chemical elements on the stellar surface has been selected
to search for an evidence of magnetic fields.

\section{Results}

193 new mean longitudinal magnetic field measurements for Ap, Bp stars and 
so-called ``non-magnetic'' stars are presented in Tables~\ref{table1} and \ref{table2}, respectively.
In the first two columns we give the HD number and another identifier.
The V magnitude and the spectral type are retrieved from the ``General Catalogue of Ap and Am stars'' by
Renson et~al.\ (1991) and in part from the SIMBAD database in case the studied stars
had no entry in the catalogue.
The modified Julian date of the middle of the exposures and
the measured mean longitudinal magnetic field $\left<B_{\mathrm l}\right>$ are presented in columns 5 and 6.
If there are several measurements for a single star, we give the reduced $\chi^2$ for 
these measurements in column 7, following:

\begin{equation}
\chi^2/n = \frac{1}{n} \sum_{i=1}^n \left( \frac{\left< B_l \right>_i}{\sigma_i} \right)^2
\end{equation}

Finally, in column 8 we identify new detections by ND and confirmed
detections by CD.
We would like to point out that all claimed detections have a significance of at
least 3\,$\sigma$, determined from the formal uncertainties we derive.
In individual cases a 3\,$\sigma$ detection could be caused by a statistical
outlier in our rather large sample of individual stars, or by slightly underestimated errors.

Because of the strong dependence of the longitudinal field on the rotational 
aspect, its usefulness to characterise actual field strength distributions is 
limited, but this can be overcome, at least in part, by repeated observations to 
sample various rotational phases, hence various aspects of the field.
Three observations per star should be the strict minimum to 
give a meaningful estimate of the intrinsic strength of the magnetic field
of a star. This estimate consists in the rms longitudinal field, 
which is computed from all $n$ measurements according to:

\begin{equation}
\overline{\left< B_l \right>} = \left( \frac{1}{n} \sum^{n}_{i=1} \left< B_l \right> ^2_i \right)^{1/2}
\label{eqn1}
\end{equation}

\begin{figure}
\resizebox{\hsize}{!}
{\includegraphics[width=0.40\textwidth]{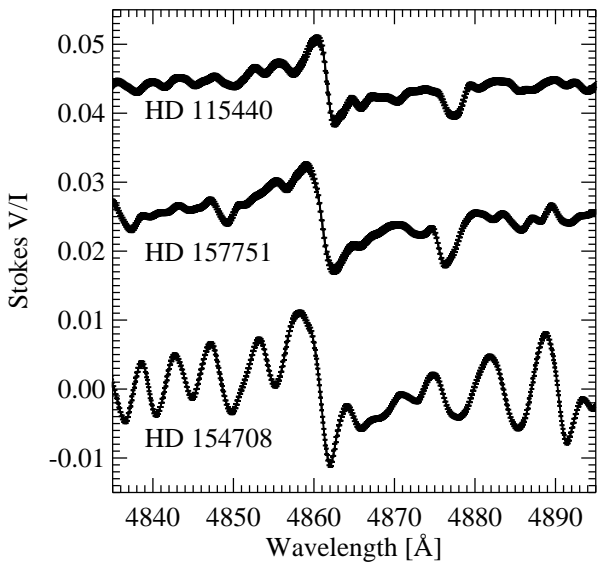}}
\caption{
V/I spectra of the newly discovered magnetic stars HD~115440, HD~157751, and HD~154708 in the 
vicinity of the H$_\beta$ line.
}
\label{fig1}
\end{figure}

While we asked for three observations per star, 
unfortunately, quite a number of stars of our program could be observed only once or
twice.
In the course of our systematical search for magnetic fields in 136 upper main sequence 
chemically peculiar stars with good Hipparcos parallaxes and in a wider range
of mass, we discovered 67 new magnetic stars.
For five other stars we could confirm earlier detections listed in the catalogue by Bychkov, 
Bychkova \& Madej (2003).
In 15 stars we confirmed our own detections a second or third time.
In Fig.~\ref{fig1} we present V/I spectra  in the vicinity of the H$_\beta $ line in
one of the most massive stars in our  sample, HD~115440, one star of intermediate mass, HD~157751,
and in the low mass star HD~154708.

In the sample of Ap and Bp stars (Table~\ref{table1}), six roAp stars show
magnetic fields well above the 3 $\sigma$ level, strongly underlying 
the close observational connection between magnetic field and pulsation:
HD~42659, HD~60435, HD~80316, HD~84041, HD~86181 and HD~154708.
The star HD~154708 is likely one of the coolest and least massive among the Ap stars and 
exhibits the second-largest mean magnetic field modulus, 24.5 kG, ever measured in an Ap star 
(Hubrig et~al.\ 2005a). Low-amplitude pulsations in HD~154708 have recently been discovered 
by Kurtz et~al.\ (2006, in preparation). 

\begin{figure}
\resizebox{\hsize}{!}
{\includegraphics[width=0.40\textwidth]{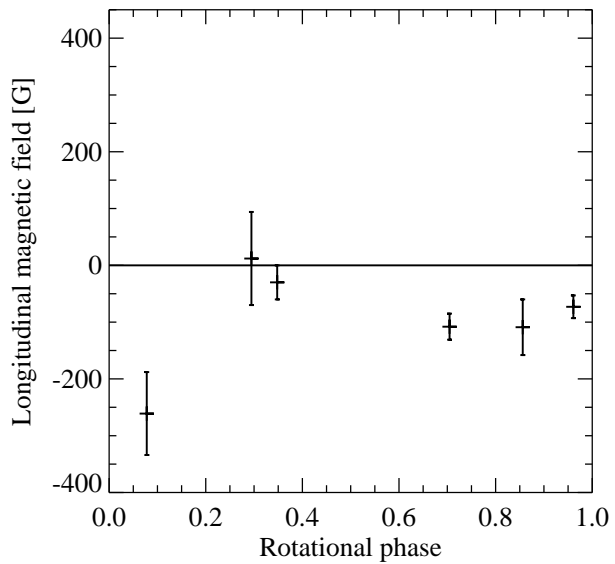}}
\caption{
Longitudinal magnetic field measurements in $\alpha$\,And as a function
of rotational phase.
}
\label{fig2}
\end{figure}

Among the 17 HgMn stars, weak magnetic fields have been detected in four stars,
HD~358 (= HR15, $\alpha$\,And), HD~65949, HD~65950, and HD~175640.
Longitudinal magnetic field measurements in $\alpha$\,And as a function
of rotational phase are presented in Fig.~\ref{fig2}.
The phases have been calculated according to the rotational ephemeris
of Adelman et~al.\ (2002).
We also detected a magnetic field at $>$3\,$\sigma$ level in 
one PGa star, HD~19400, and in the normal B-type star HD~179761.
Five years ago we already showed
evidence for a relative magnetic intensification of Fe\,II lines
produced by different magnetic desaturations induced by different
Zeeman-split components in HD~179761 (Hubrig \& Castelli 2001).
As the relative intensification is roughly correlated
with the strength of the magnetic field, it is a powerful tool for detecting
magnetic fields which have a complex structure and are difficult to detect by
polarization measurements.

Weak magnetic fields have also been discovered in four SPB stars:
HD~53921, HD~74560, HD~85953, and HD~215573.
There have been only a few isolated attempts to determine magnetic
fields in SPB stars. Neiner et~al.\ (2003) searched for a magnetic field in
the B2IV star $\zeta$\,Cas, which lies in the region of the H-R diagram that belongs
both to the SPB and the $\beta$\,Cep instability strip. Using time-resolved
spectropolarimetric observations with the Musicos echelle spectropolarimeter
at the 2\,m Telescope Bernard Lyot they obtained 
clear Zeeman signatures indicative of the presence of a magnetic field over the 
rotational period of 5.4~d. This star was the first known magnetic SPB star.
However, the role that magnetic fields
play in the understanding of pulsational properties of SPB stars is still
unclear, and further observations are needed to look for possible relations
between magnetic field and pulsation patterns.

Normal B, HgMn, PGa, and SPB stars
are usually regarded as non-magnetic stars.
However, the intriguing discovery of mean
longitudinal magnetic fields of the order of a few hundred Gauss in a sample
of so-called ``non-magnetic'' stars rises a fundamental question about the
possible ubiquitous presence of a magnetic field in upper main sequence stars.
The structure of the field in these stars must be, however,
sufficiently tangled so that it does not produce a strong net observable
circular polarization signature.

In this paper, we presented results of our comprehensive study of magnetic fields
in 136 upper main sequence stars. The magnetic field determination method
is based on circular polarized FORS\,1 spectra and shows the excellent potential of FORS\,1 
for measuring magnetic fields.
The preliminary results of our analysis of the evolutionary state
of magnetic chemically peculiar stars based on the smaller sample of stars measured with FORS\,1 have 
been presented in the last years as meeting contributions (Hubrig et~al.\ 2005b;
Hubrig et~al.\ 2006).
In Paper II we will present the complete analysis based 
on a large sample of magnetic stars.

\acknowledgements

We would like to thank Thomas Szeifert for providing the
FORS\,1 spectra extraction routines.
This research made use of the SIMBAD database,
operated at the CDS, Strasbourg, France.

\begin{appendix}
\section{List of magnetic field measurements}

\begin{table*}
\caption{
The mean longitudinal field measurements for our sample of Ap and Bp stars observed with FORS\,1
in the frame of our ESO service programs 71.D-0308, 072.D-0377, 073.D-0464, and 075.D-0295.
In the first two columns we give the HD number and another identifier.
The V magnitude and the spectral type are retrieved from the ``General Catalogue of Ap and Am stars'' by
Renson et~al.\ (1991) and in part from the SIMBAD database in case the studied stars
had no entry in the catalogue.
The modified Julian date of the middle of the exposures and 
the measured mean longitudinal magnetic field $\left<B_{\mathrm l}\right>$
are presented in columns 5 and 6.
If there are several measurements for a single star, we give the reduced $\chi^2$ for 
all measurements in column 7.
Finally, in column 8 we identify new detections by ND
and confirmed detections by CD (see text).
}
\label{table1}
\begin{center}
\begin{tabular}{rlrlcrrc}
\hline
\multicolumn{1}{c}{HD} &
\multicolumn{1}{c}{Other} &
\multicolumn{1}{c}{$V$} &
\multicolumn{1}{c}{Sp.\ Type} &
\multicolumn{1}{c}{MJD} &
\multicolumn{1}{c}{$\left<B_{\mathrm l}\right>$} &
\multicolumn{1}{c}{$\chi^2/n$} &
\multicolumn{1}{c}{Comment} \\
\multicolumn{1}{c}{} &
\multicolumn{1}{c}{identifier} &
\multicolumn{1}{c}{} &
\multicolumn{1}{c}{} &
\multicolumn{1}{c}{} &
\multicolumn{1}{c}{[G]} &
\multicolumn{1}{c}{} &
\multicolumn{1}{c}{} \\
\hline
1048 & HR49 & 6.2 & A1 Si & 52910.103 & 244$\pm$ 74 &  & ND \\
 &  &  &  & 53199.406 & $-$69$\pm$ 54 &  &   \\
 &  &  &  & 53215.382 & 84$\pm$ 53 &  &   \\
 &  &  &  & 53216.405 & 56$\pm$ 45 & 4.1 &   \\
3326 & HR151 & 6.1 & A6 Sr & 52908.190 & 49$\pm$ 49 &  &   \\
3980 & HR183 & 5.7 & A7 Sr Eu Cr & 53559.410 & 1210$\pm$ 32 &  & CD \\
 &  &  &  & 53624.076 & 395$\pm$ 26 &  & CD \\
 &  &  &  & 53630.232 & 452$\pm$ 15 & 856.2 & CD \\
8783 & CP$-$72~~~98 & 7.8 & A2 Sr Eu Cr & 52852.358 & 26$\pm$ 92 &  &   \\
10840 & CP$-$61~~139 & 6.8 & B9 Si & 53184.331 & $-$161$\pm$131 &  &   \\
19712 & BD$-$02~~563 & 7.3 & A0 Cr Eu & 52905.384 & $-$963$\pm$ 77 &  & ND \\
 &  &  &  & 52999.025 & 550$\pm$ 42 & 163.9 & CD \\
19918 & CP$-$82~~~54 & 9.4 & A5 Sr Eu Cr & 52908.210 & $-$625$\pm$ 87 &  & CD \\
22374 & BD+22~~518 & 6.7 & A1 Cr Sr Si & 52999.039 & 31$\pm$ 57 &  &   \\
 &  &  &  & 53216.383 & $-$72$\pm$ 59 & 0.9 &   \\
22488 & CP$-$67~~236 & 7.7 & A3 Sr Eu Cr & 53087.014 & 102$\pm$ 53 &  &   \\
23207 & BD$-$19~~732 & 7.5 & A2 Sr Eu & 53215.361 & 259$\pm$ 92 &  &   \\
 &  &  &  & 53218.338 & 411$\pm$ 94 & 13.5 & ND \\
24188 & CP$-$72~~262 & 6.3 & A0 Si & 53087.032 & 404$\pm$ 55 &  & ND \\
30612 & HR1541 & 5.5 & B9 Si & 53087.046 & 10$\pm$ 51 &  &   \\
34797 & HR1754 & 6.5 & B8 He-weak Si & 52999.066 & 713$\pm$ 54 &  & ND \\
34798 & HR1753 & 6.5 & B8 He-weak Si & 52999.055 & 56$\pm$ 82 &  &   \\
42659 & BD$-$15~1299 & 6.7 & A3 Sr Cr Eu & 52999.119 & 392$\pm$ 72 &  & ND \\
55522 & HR2718 & 5.9 & B2IV/V & 52999.190 & 38$\pm$ 73 &  &   \\
 &  &  &  & 52999.227 & 39$\pm$234 &  &   \\
 &  &  &  & 53000.053 & 873$\pm$ 66 & 58.4 & ND \\
56350 & CP$-$53~1284 & 6.7 & A0 Eu Cr Sr & 52999.239 & 736$\pm$125 &  & ND \\
56455 & HR2761 & 5.7 & A0 Si & 52999.251 & $-$119$\pm$ 70 &  &   \\
58448 & CP$-$61~~814 & 7.1 & B8 Si & 52999.265 & $-$331$\pm$ 66 &  & ND \\
60435 & CP$-$57~1246 & 8.9 & A3 Sr Eu & 53000.072 & $-$296$\pm$ 52 &  & ND \\
63401 & HR3032 & 6.3 & B9 Si & 53002.053 & $-$236$\pm$ 70 &  & ND \\
 &  &  &  & 53004.228 & $-$656$\pm$ 75 & 43.9 & CD \\
68826 & CO$-$48~3586 & 9.3 & B9 Si & 53454.077 & 80$\pm$ 64 &  &   \\
69144 & HR3244 & 5.1 & B2.5IV & 52989.350 & 41$\pm$ 63 &  &   \\
74168 & CO$-$51~3141 & 7.5 & B9 Si & 53002.111 & $-$437$\pm$101 &  & ND \\
74196 & HR3448 & 5.6 & B7 He-weak  & 52906.388 & 254$\pm$118 &  &   \\
75989 & CO$-$40~4685 & 6.5 & B9 Si & 52992.341 & $-$368$\pm$107 &  & ND \\
 &  &  &  & 53004.286 & $-$409$\pm$105 & 13.5 & CD \\
80316 & BD$-$19~2674 & 7.8 & A3 Sr Eu & 52992.357 & $-$183$\pm$ 38 &  & ND \\
83625 & CP$-$53~2664 & 6.9 & A0 Si Sr & 53008.325 & $-$1208$\pm$ 64 &  & ND \\
84041 & CO$-$28~7536A & 9.4 & A5 Sr Eu & 53002.170 & 479$\pm$ 72 &  & ND \\
86181 & CP$-$58~1700 & 9.4 & F0 Sr & 53002.201 & 404$\pm$ 94 &  & ND \\
86199 & CP$-$56~2646 & 6.7 & B9 Si & 53003.345 & $-$921$\pm$ 67 &  & ND \\
88158 & CP$-$61~1479 & 6.5 & B8 Si & 53008.338 & $-$8$\pm$ 75 &  &   \\
88385 & CP$-$56~2919 & 8.1 & A0 Cr Eu Si & 53010.181 & $-$1054$\pm$ 65 &  & ND \\
89103 & CO$-$48~5469 & 7.8 & B9 Si & 53010.202 & $-$2303$\pm$ 48 &  & ND \\
89385 & CP$-$53~3579 & 8.4 & B9 Cr Eu Si & 53010.218 & $-$255$\pm$ 61 &  & ND \\
90264 & HR4089 & 5.0 & B8 He-weak  & 52824.019 & 114$\pm$108 &  &   \\
91239 & CO$-$41~5923 & 7.4 & B9 Eu Cr Si & 53118.059 & $-$33$\pm$ 89 &  &   \\
92106 & CP$-$80~~468 & 7.8 & A0 Sr Eu Cr & 53010.239 & $-$258$\pm$ 71 &  & ND \\
 &  &  &  & 53118.080 & $-$243$\pm$102 & 9.4 &   \\
92385 & CP$-$64~1374 & 6.7 & B9 Si & 53008.369 & $-$623$\pm$100 &  & ND \\
 &  &  &  & 53020.332 & $-$165$\pm$ 59 & 23.3 &   \\
\hline
\end{tabular}
\end{center}
\end{table*}

\addtocounter{table}{-1}
\begin{table*}
\caption{Continued.}
\begin{center}
\begin{tabular}{rlrlcrrc}
\hline
\multicolumn{1}{c}{HD} &
\multicolumn{1}{c}{Other} &
\multicolumn{1}{c}{$V$} &
\multicolumn{1}{c}{Sp.\ Type} &
\multicolumn{1}{c}{MJD} &
\multicolumn{1}{c}{$\left<B_{\mathrm l}\right>$} &
\multicolumn{1}{c}{$\chi^2/n$} &
\multicolumn{1}{c}{Comment} \\
\multicolumn{1}{c}{} &
\multicolumn{1}{c}{identifier} &
\multicolumn{1}{c}{} &
\multicolumn{1}{c}{} &
\multicolumn{1}{c}{} &
\multicolumn{1}{c}{[G]} &
\multicolumn{1}{c}{} &
\multicolumn{1}{c}{} \\
\hline
92499 & CO$-$42~6407 & 8.9 & A2 Sr Eu Cr & 53010.255 & $-$964$\pm$172 &  & ND \\
 &  &  &  & 53011.212 & $-$1255$\pm$ 69 &  & CD \\
 &  &  &  & 53118.095 & $-$1191$\pm$148 & 142.3 & CD \\
93030 & HR4199 & 2.7 & B0 Si N P & 53012.231 & $-$205$\pm$137 &  &   \\
96451 & CP$-$74~~771 & 6.9 & A0 Sr & 53074.346 & 108$\pm$ 70 &  &   \\
98340 & CP$-$58~3433 & 7.1 & B9 Si & 53074.362 & 977$\pm$ 73 &  & ND \\
99563 & BD$-$08~3173A & 8.5 & F0 Sr & 53012.247 & $-$235$\pm$ 73 &  & CD \\
 &  &  &  & 53015.225 & $-$670$\pm$ 84 & 37.0 & CD \\
105379 & CO$-$30~9691 & 8.0 & A0 Sr Cr & 53011.250 & $-$283$\pm$ 74 &  & ND \\
105382 & HR4618 & 4.4 & B6IIIe & 53011.195 & $-$923$\pm$ 86 &  & ND \\
 &  &  &  & 53015.247 & $-$431$\pm$109 & 65.4 & CD \\
105770 & CP$-$83~~444 & 7.4 & B9 Si & 53011.233 & 160$\pm$ 58 &  &   \\
 &  &  &  & 53120.145 & 254$\pm$ 83 & 8.5 & ND \\
105999 & CP$-$62~2619 & 7.4 & F1 Sr Cr & 53011.270 & $-$247$\pm$ 58 &  & ND \\
107696 & HR4706 & 5.4 & B8 Cr & 52824.030 & $-$9$\pm$ 90 &  &   \\
 &  &  &  & 53074.375 & $-$134$\pm$145 & 0.4 &   \\
108945 & HR4766 & 5.5 & A3 Sr & 53015.335 & $-$347$\pm$ 51 &  & ND \\
114365 & HR4965 & 6.1 & A0 Si & 52824.043 & $-$24$\pm$ 57 &  &   \\
115226 & CP$-$72~1373 & 8.5 & A3 Sr & 53074.392 & 820$\pm$139 &  & ND \\
 &  &  &  & 53086.299 & 654$\pm$ 66 & 66.5 & CD \\
115440 & CP$-$75~~859 & 8.2 & B9 Si & 53077.215 & 3120$\pm$ 73 &  & ND \\
116890 & HR5066 & 6.2 & B9 Si & 52824.055 & $-$119$\pm$ 62 &  &   \\
117025 & HR5069 & 6.1 & A2 Sr Eu Cr & 52824.067 & 455$\pm$ 73 &  & ND \\
 &  &  &  & 53120.164 & 416$\pm$ 94 & 29.2 & CD \\
118913 & CP$-$68~1981 & 7.7 & A0 Eu Cr Sr & 52824.081 & $-$385$\pm$ 71 &  & ND \\
 &  &  &  & 53120.181 & $-$544$\pm$ 75 & 41.0 & CD \\
119308 & CO$-$34~9094 & 7.8 & B9 Sr Cr Eu & 53120.204 & $-$325$\pm$ 73 &  & ND \\
122970 & BD+06~2827 & 8.3 & F0p & 53015.350 & 352$\pm$101 &  & CD \\
125630 & CP$-$66~2519 & 6.8 & A2 Si Cr Sr & 52824.107 & 659$\pm$ 54 &  & ND \\
 &  &  &  & 53120.221 & 9$\pm$ 63 & 74.5 &   \\
127453 & CP$-$68~2132 & 7.4 & B8 Si & 52824.121 & $-$360$\pm$ 69 &  & ND \\
127575 & CP$-$68~2135 & 7.7 & B9 Si & 53079.388 & 807$\pm$ 72 &  & ND \\
128775 & CO$-$45~9337 & 6.6 & B9 Si & 53120.236 & $-$340$\pm$ 61 &  & ND \\
128974 & HR5466 & 5.7 & A0 Si & 52824.144 & 40$\pm$ 45 &  &   \\
129899 & CP$-$76~~894 & 6.4 & A0 Si & 53120.295 & 402$\pm$ 48 &  & ND \\
130158 & HR5514 & 5.6 & B9 Si & 52824.176 & 28$\pm$ 44 &  &   \\
 &  &  &  & 53116.312 & 51$\pm$ 43 & 0.9 &   \\
130557 & HR5522 & 6.1 & B9 Si Cr & 52853.058 & $-$30$\pm$ 74 &  &   \\
 &  &  &  & 53144.267 & 100$\pm$ 50 & 2.1 &   \\
131120 & HR5543 & 5.0 & B7 He-weak  & 52824.158 & $-$228$\pm$110 &  &   \\
 &  &  &  & 53020.353 & $-$137$\pm$ 74 &  &   \\
 &  &  &  & 53030.366 & 63$\pm$ 69 & 2.9 &   \\
132322 & CP$-$63~3473 & 7.4 & A7 Sr Cr Eu & 53111.311 & 357$\pm$ 51 &  & ND \\
133792 & HR5623 & 6.3 & A0 Sr Cr & 52853.070 & $-$55$\pm$116 &  &   \\
 &  &  &  & 53120.312 & 68$\pm$ 47 & 1.2 &   \\
134305 & BD+13~2899 & 7.2 & A6 Sr Eu Cr & 53144.301 & 117$\pm$ 68 &  &   \\
136933 & HR5719 & 5.4 & A0 Si & 52823.223 & 56$\pm$ 68 &  &   \\
138758 & CP$-$74~1451 & 7.9 & B9 Si & 53086.328 & 415$\pm$ 47 &  & ND \\
138764 & HR5780 & 5.2 & B6 Si & 52904.016 & 146$\pm$ 57 &  &   \\
138769 & HR5781 & 4.5 & B3IVp & 52904.027 & $-$16$\pm$ 58 &  &   \\
 &  &  &  & 52908.022 & $-$260$\pm$ 84 & 4.8 & ND \\
145102 & CO$-$2611240 & 6.6 & B9 Si & 52763.315 & $-$48$\pm$ 75 &  &   \\
\hline
\end{tabular}
\end{center}
\end{table*}

\addtocounter{table}{-1}
\begin{table*}
\caption{Continued.}
\begin{center}
\begin{tabular}{rlrlcrrc}
\hline
\multicolumn{1}{c}{HD} &
\multicolumn{1}{c}{Other} &
\multicolumn{1}{c}{$V$} &
\multicolumn{1}{c}{Sp.\ Type} &
\multicolumn{1}{c}{MJD} &
\multicolumn{1}{c}{$\left<B_{\mathrm l}\right>$} &
\multicolumn{1}{c}{$\chi^2/n$} &
\multicolumn{1}{c}{Comment} \\
\multicolumn{1}{c}{} &
\multicolumn{1}{c}{identifier} &
\multicolumn{1}{c}{} &
\multicolumn{1}{c}{} &
\multicolumn{1}{c}{} &
\multicolumn{1}{c}{[G]} &
\multicolumn{1}{c}{} &
\multicolumn{1}{c}{} \\
\hline
147869 & HR6111 & 5.8 & A1 Sr & 52763.327 & 28$\pm$ 62 &  &   \\
 &  &  &  & 53144.318 & $-$68$\pm$ 47 & 1.1 &   \\
148112 & HR6117 & 4.6 & A0 Cr Eu & 52763.338 & $-$62$\pm$ 53 &  &   \\
148898 & HR6153 & 4.4 & A6 Sr Cr Eu & 52763.349 & 175$\pm$ 70 &  &   \\
149764 & CO$-$3811087 & 6.9 & A0 Si & 52763.374 & $-$1213$\pm$ 70 &  & ND \\
 &  &  &  & 53120.325 & 20$\pm$ 68 &  &   \\
 &  &  &  & 53120.335 & 30$\pm$ 57 & 100.2 &   \\
149822 & HR6176 & 6.4 & B9 Si Cr & 52763.361 & $-$645$\pm$ 54 &  & ND \\
150549 & HR6204 & 5.1 & A0 Si & 52763.386 & $-$187$\pm$ 51 &  & ND \\
 &  &  &  & 53116.386 & $-$228$\pm$ 59 &  & CD \\
 &  &  &  & 53120.350 & $-$110$\pm$ 52 & 11.0 &   \\
151525 & HR6234 & 5.2 & B9 Eu Cr & 52733.395 & $-$14$\pm$ 60 &  &   \\
 &  &  &  & 52763.397 & 186$\pm$ 61 & 4.7 & CD \\
154708 & CP$-$57~8336 & 8.8 & A2 Sr Eu Cr & 53120.376 & 7530$\pm$ 54 &  & ND \\
 &  &  &  & 53487.302 & 5764$\pm$ 25 &  & CD \\
 &  &  &  & 53519.344 & 5819$\pm$ 52 & 28375.0 & CD \\
157751 & CO$-$3312069 & 7.6 & B9 Si Cr & 52793.271 & 4063$\pm$ 54 &  & ND \\
 &  &  &  & 53116.404 & 3968$\pm$ 55 & 5433.1 & CD \\
160468 & CP$-$68~2936 & 7.3 & F2 Sr Cr & 53116.362 & 63$\pm$ 63 &  &   \\
 &  &  &  & 53134.319 & $-$205$\pm$101 & 2.6 &   \\
161277 & CO$-$3911816 & 7.1 & B9 Si & 53134.339 & 1$\pm$ 42 &  &   \\
166469 & HR6802 & 6.5 & A0 Si Cr Sr & 52793.287 & $-$15$\pm$ 55 &  &   \\
 &  &  &  & 52793.295 & $-$133$\pm$ 60 &  &   \\
 &  &  &  & 53136.274 & $-$42$\pm$ 52 & 1.9 &   \\
168856 & BD$-$07~4589 & 7.0 & B9 Si & 53144.343 & $-$608$\pm$ 47 &  & ND \\
171184 & BD$-$14~5110 & 8.0 & A0 Si & 52880.042 & 379$\pm$ 58 &  & ND \\
 &  &  &  & 53144.368 & $-$54$\pm$ 48 & 22.0 &   \\
171279 & BD$-$07~4623 & 7.3 & A0 Sr Cr Eu & 53144.393 & $-$45$\pm$ 46 &  &   \\
172032 & BD$-$16~4963 & 7.7 & A9 Sr Cr & 53151.105 & $-$8$\pm$ 67 &  &   \\
172690 & CP$-$84~~587 & 7.5 & A0 Si Sr Cr & 52793.314 & $-$225$\pm$ 71 &  & ND \\
 &  &  &  & 53134.368 & 230$\pm$ 60 & 12.4 & CD \\
175744 & HR7147 & 6.6 & B9 Si & 52901.019 & 147$\pm$ 53 &  &   \\
176196 & CP$-$74~1739 & 7.5 & B9 Eu Cr & 52793.329 & 258$\pm$ 69 &  & ND \\
 &  &  &  & 53134.389 & 174$\pm$ 58 & 11.5 & CD \\
183806 & HR7416 & 5.6 & A0 Cr Eu Sr & 52793.345 & $-$229$\pm$ 45 &  & ND \\
 &  &  &  & 53120.424 & 172$\pm$ 43 & 20.9 & CD \\
186117 & CP$-$73~2061 & 7.3 & A0 Sr Cr Eu & 53134.413 & $-$52$\pm$ 55 &  &   \\
 &  &  &  & 53140.329 & $-$36$\pm$ 54 & 0.7 &   \\
192674 & CO$-$5112473 & 7.5 & B9 Cr Eu Sr & 53137.362 & $-$30$\pm$ 45 &  &   \\
199180 & BD+16~4401 & 7.7 & A0 Si Cr & 52822.344 & $-$215$\pm$ 74 &  &   \\
199728 & HR8033 & 6.2 & B9 Si & 52822.357 & $-$254$\pm$ 60 &  & ND \\
201018 & CO$-$3714125 & 8.6 & A2 Cr Eu & 53151.371 & 494$\pm$153 &  & ND \\
202627 & HR8135 & 4.7 & A1 Si & 52793.374 & $-$118$\pm$ 57 &  &   \\
206653 & CP$-$68~3444 & 7.2 & B9 Si & 52793.394 & 125$\pm$ 83 &  &   \\
212385 & CO$-$3914697 & 6.8 & A3 Sr Eu Cr & 52822.413 & 145$\pm$ 59 &  &   \\
 &  &  &  & 53184.297 & 541$\pm$ 60 & 43.7 & ND \\
221760 & HR8949 & 4.7 & A2 Sr Cr Eu & 52793.415 & $-$103$\pm$ 80 &  &   \\
 &  &  &  & 53184.314 & 16$\pm$ 85 & 0.8 &   \\
223640 & HR9031 & 5.2 & B9 Si Sr Cr & 52822.428 & $-$74$\pm$ 51 &  &   \\
\hline
\end{tabular}
\end{center}
\end{table*}

\begin{table*}
\caption{
The mean longitudinal field measurements for our sample of so-called ``non-magnetic'' stars
observed with FORS\,1 in the frame of our ESO service programs 71.D-0308, 072.D-0377, 073.D-0464, and 075.D-0295. 
In the first two columns we give the HD number and another identifier.
The V magnitude and the spectral type are retrieved from the ``General Catalogue of Ap and Am stars'' by
Renson et~al.\ (1991) and in part from the SIMBAD database in case the studied stars
had no entry in the catalogue.
The modified Julian date of the middle of the exposures and 
the measured mean longitudinal magnetic field $\left<B_{\mathrm l}\right>$
are presented in columns 5 and 6.
If there are several measurements for a single star, we give the reduced $\chi^2$ for 
all measurements in column 7.
In column 8 we identify new detections by ND and confirmed detections by CD (see text).
}
\label{table2}
\begin{center}
\begin{tabular}{rlrlcrrc}
\hline
\multicolumn{1}{c}{HD} &
\multicolumn{1}{c}{Other} &
\multicolumn{1}{c}{$V$} &
\multicolumn{1}{c}{Sp.\ Type} &
\multicolumn{1}{c}{MJD} &
\multicolumn{1}{c}{$\left<B_{\mathrm l}\right>$} &
\multicolumn{1}{c}{$\chi^2/n$} &
\multicolumn{1}{c}{Comment} \\
\multicolumn{1}{c}{} &
\multicolumn{1}{c}{identifier} &
\multicolumn{1}{c}{} &
\multicolumn{1}{c}{} &
\multicolumn{1}{c}{} &
\multicolumn{1}{c}{[G]} &
\multicolumn{1}{c}{} &
\multicolumn{1}{c}{} \\
\hline
\hline
\multicolumn{8}{c}{HgMn stars} \\
\hline
358 & HR15 & 2.1 & B9 Mn Hg & 52910.092 & $-$261$\pm$ 73 &  & ND \\
 &  &  &  & 52963.020 & 12$\pm$ 82 &  &   \\
 &  &  &  & 53519.448 & $-$109$\pm$ 49 &  &   \\
 &  &  &  & 53629.286 & $-$73$\pm$ 20 &  & CD \\
 &  &  &  & 53630.208 & $-$30$\pm$ 30 &  &   \\
 &  &  &  & 53638.205 & $-$108$\pm$ 23 & 9.0 & CD \\
23408 & HR1149 & 3.9 & B7 He-weak Mn & 52963.156 & $-$83$\pm$ 46 &  &   \\
23950 & HR1185 & 6.1 & B9 Mn Hg Si & 53215.403 & 62$\pm$ 79 &  &   \\
 &  &  &  & 53216.418 & 50$\pm$ 55 & 0.7 &   \\
49606 & HR2519 & 5.8 & B8 Mn Hg Si & 52946.354 & $-$11$\pm$ 71 &  &   \\
53929 & HR2676 & 6.1 & B9 Mn Hg & 52992.306 & $-$178$\pm$129 &  &   \\
 &  &  &  & 53004.210 & $-$248$\pm$108 & 3.6 &   \\
63975 & HR3059 & 5.1 & B8 Mn Hg & 52992.278 & 95$\pm$ 74 &  &   \\
65949 & CP$-$60~~966 & 8.4 & B9 Hg & 53002.082 & $-$290$\pm$ 62 &  & ND \\
65950 & CP$-$60~~967 & 6.9 & B9 Mn Hg & 53002.067 & $-$179$\pm$ 53 &  & ND \\
71066 & HR3302 & 5.6 & A0 Si Mn & 53002.098 & 1$\pm$ 56 &  &   \\
87752 & CP$-$59~1843 & 9.8 & B9 Hg Mn & 53008.304 & $-$151$\pm$100 &  &   \\
155379 & HR6386 & 6.5 & A0 Hg Y & 52763.410 & 46$\pm$ 52 &  &   \\
 &  &  &  & 53137.393 & $-$65$\pm$ 46 & 1.4 &   \\
175640 & HR7143 & 6.2 & A0 Hg Mn Y & 52901.043 & 207$\pm$ 65 &  & ND \\
186122 & HR7493 & 6.3 & B9 Mn Hg & 52822.312 & 215$\pm$ 76 &  &   \\
194783 & HR7817 & 6.1 & B9 Hg Mn & 52793.361 & $-$43$\pm$ 53 &  &   \\
202149 & HR8118 & 6.7 & B9 Hg & 53137.413 & 43$\pm$ 39 &  &   \\
202671 & HR8137 & 5.4 & B7 He-weak Mn & 53151.411 & $-$109$\pm$ 57 &  &   \\
221507 & HR8937 & 4.4 & B9 Mn Hg & 52900.092 & $-$61$\pm$ 36 &  &   \\
\hline
\multicolumn{8}{c}{SPB stars} \\
\hline
24587 & HR1213 & 4.6 & B6 & 52971.071 & $-$120$\pm$ 68 &  &   \\
26326 & HR1288 & 5.4 & B5IV & 52909.389 & 119$\pm$ 80 &  &   \\
53921 & HR2674 & 5.6 & B9IV & 52999.137 & $-$294$\pm$ 63 &  & ND \\
74195 & HR3447 & 3.6 & B3IV & 53002.123 & $-$277$\pm$108 &  &   \\
74560 & HR3467 & 4.8 & B3 Mg Si & 53002.141 & $-$199$\pm$ 61 &  & ND \\
85953 & HR3924 & 5.9 & B2III & 53002.152 & $-$131$\pm$ 42 &  & ND \\
92287 & HR4173 & 5.9 & B3IV & 53008.352 & $-$10$\pm$ 57 &  &   \\
123515 & HR5296 & 6.0 & B8 Si & 52824.093 & $-$59$\pm$ 50 &  &   \\
215573 & HR8663 & 5.3 & B6IV & 52900.080 & 165$\pm$ 53 &  & ND \\
\hline
\multicolumn{8}{c}{PGa stars} \\
\hline
19400 & HR939 & 5.5 & B8 He-weak  & 52852.371 & 217$\pm$ 65 &  & ND \\
120709 & HR5210 & 4.6 & B5 He-weak P & 53015.323 & 79$\pm$ 76 &  &   \\
\hline
\multicolumn{8}{c}{Normal B stars} \\
\hline
91375 & HR4138 & 4.7 & A2 & 53116.028 & $-$58$\pm$ 56 &  &   \\
179761 & HR7287 & 5.1 & B8  & 52822.280 & $-$267$\pm$ 68 &  & ND \\
209459 & HR8404 & 5.8 & B9  & 52822.381 & $-$144$\pm$ 60 &  &   \\
\hline
\end{tabular}
\end{center}
\end{table*}

\end{appendix}

\end{document}